\documentclass[11pt,a4paper]{article}
\usepackage[margin=2.4cm]{geometry}
\usepackage[T1]{fontenc}
\usepackage[utf8]{inputenc}
\usepackage{lmodern,microtype}
\usepackage{amsmath,amssymb,amsthm,mathtools,bm,booktabs,mathrsfs}
\usepackage{hyperref}
\hypersetup{colorlinks=true,linkcolor=blue,citecolor=blue,urlcolor=blue,pdftitle={Dynamical Gauss-Bonnet Sum Rules and Global Consistency in Five-Dimensional Braneworld Cosmology},pdfauthor={Paulo Michel Longo Tavares da Silva}}
\usepackage{enumitem}
\newtheorem{theorem}{Theorem}
\newtheorem{corollary}{Corollary}
\newcommand{\aGB}{\alpha_{\mathrm{GB}}}
\newcommand{\kap}{\kappa_5^2}
\newcommand{\dd}{\mathrm d}
\newcommand{\cG}{\mathcal G}
\newcommand{\ointy}{\oint_{S^1}\!\dd y\,}
\title{\bfseries Dynamical Gauss--Bonnet Sum Rules and Global Consistency in Five-Dimensional Braneworld Cosmology}
\author{Paulo Michel Longo Tavares da Silva\\[3pt]\small Faculty of Technology of Botucatu (FATEC Botucatu)\\\small Paula Souza State Center for Technological Education (CEETEPS)\\\small Botucatu, S\~ao Paulo, Brazil\\\small \texttt{paulo.silva19@cps.sp.gov.br}}
\date{}
\begin{document}
\maketitle
\begin{abstract}
We derive an exact dynamical consistency relation for a smooth, periodically compactified five-dimensional Einstein--Gauss--Bonnet cosmology sourced by a real, minimally coupled canonical scalar field. For a spatially flat Friedmann--Lema\^itre--Robertson--Walker slicing and a generally time-dependent warp factor, a contraction of the bulk equations eliminates both the scalar potential and the bulk cosmological constant. A direct orthonormal-curvature reconstruction gives the independent diagonal and mixed field equations and verifies the contracted Gauss--Bonnet identity. The integrated relation separates an exact total derivative from the dynamically active Lovelock sector. Importantly, satisfying the integrated sum rule alone does not imply a solution of the complete system. We prove that, for positive Gauss--Bonnet coupling, a smooth periodic \emph{time-independent} warp factor and a canonical scalar exclude all exact expanding power-law scale factors $a(t)\propto t^p$ with $p>0$. This strengthens the previously established de Sitter sign obstruction and removes an earlier interpretation of $p=1/2$ as an admissible sign-compatible physical branch. For time-dependent warping, we reformulate bouncing conditions using proper time and the physical scale factor on each constant-$y$ hypersurface. We compare the common $w=-1$ geometric member with the perturbative weighted construction of Dias, Hoff da Silva, and da Rocha. At the level of the unexpanded Ricci-trace rearrangement, the curvature-only tensor contribution differs from the full Lanczos contraction by a definite scalar-curvature term. This distinction, and the additional independent field equations, delimit the physical conclusions that may be drawn from a necessary integrated condition.
\end{abstract}
\noindent\textbf{Keywords:} Einstein--Gauss--Bonnet gravity; Lovelock theory; braneworld cosmology; compact extra dimension; consistency conditions; no-go theorem; physical cosmological bounce.

\section{Introduction}
Global braneworld consistency conditions provide necessary constraints that need not be evident from a restricted set of local equations. Gibbons, Kallosh, and Linde~\cite{GKL2001} identified an obstruction to smooth scalar-supported periodic compactifications in five-dimensional Einstein gravity. The general weighted construction was developed by Leblond, Myers, and Winters~\cite{LMW2001}. Including the Gauss--Bonnet density in five dimensions adds the lowest nontrivial higher-curvature Lovelock contribution~\cite{Lovelock1971,Padmanabhan2013}.

The braneworld literature developed from the Randall--Sundrum geometries~\cite{RS1999a,RS1999b} to smooth scalar-supported walls and gravitational localization~\cite{DeWolfe2000,Gremm2000,Csaki2000}. Einstein--Gauss--Bonnet gravity and its cosmological branches have been investigated in higher-curvature theory~\cite{BD1985,Abdesselam2002,Charmousis2002}, junction conditions~\cite{Davis2003,Gravanis2003}, and perturbative or cosmological applications~\cite{Dufaux2004,Maeda2007}. Time-dependent warped geometries and FRW braneworlds provide complementary contexts~\cite{Guha2010,Guha2018,Michel2017}; other extensions include variable brane tensions~\cite{Abdalla2009} and nonminimal scalar--Gauss--Bonnet couplings~\cite{Silva2026}. These constructions have distinct matter and compactification assumptions and should not be conflated with the system considered below.

Previous weighted Gauss--Bonnet constructions, including Dias, Hoff da Silva, and da Rocha~\cite{Dias2015}, focus on static warping and, in key applications, a weak higher-curvature approximation. We instead start from the exact five-dimensional field equations and permit a time-dependent warp factor. The key conceptual distinction is that a global integral is necessary but not sufficient for a scalar-supported warped geometry: the independent field equations may remove apparently sign-compatible sectors.

The analysis has three physical goals: (i) derive and apply the independent local constraints to power-law expansion; (ii) define a bounce through induced proper time rather than the coordinate quantity $H=\dot a/a$ alone; and (iii) compare the assumptions and level of approximation with the $w=-1$ weighted condition of Ref.~\cite{Dias2015}.

\section{Five-dimensional Einstein--Gauss--Bonnet--scalar system}
The action is
\begin{equation}
S=\int \dd^5x\,\sqrt{-g}\left\{\frac{R-2\Lambda_5+\aGB\cG}{2\kap}-\frac{1}{2}g^{AB}\partial_A\phi\,\partial_B\phi-V(\phi)\right\},
\label{eq:action}
\end{equation}
where $\cG=R^2-4R_{AB}R^{AB}+R_{ABCD}R^{ABCD}$. With the conventions of the submitted manuscript,
\begin{align}
G_{AB}+\Lambda_5g_{AB}+\aGB\mathcal H_{AB}&=\kap T_{AB},\label{eq:field}\\
T_{AB}&=\partial_A\phi\partial_B\phi-g_{AB}\left[\tfrac12(\nabla\phi)^2+V(\phi)\right],\\
\mathcal H_{AB}&=2\bigl(RR_{AB}-2R_{AC}R^C{}_B-2R^{CD}R_{ACBD}\bigr)\notag\\
&\qquad+2R_A{}^{CDE}R_{BCDE}-\tfrac12g_{AB}\cG.\label{eq:lanczos}
\end{align}
The scalar equation is $\square_5\phi-V_{,\phi}=0$ and the Lovelock tensor obeys $\nabla^A\mathcal H_{AB}=0$.

We use the geometry
\begin{equation}
\dd s^2=e^{2A(t,y)}[-\dd t^2+a^2(t)\delta_{ij}\dd x^i\dd x^j]+\dd y^2,
\quad H=\frac{\dot a}{a},\quad X=H+\dot A,
\label{eq:metric}
\end{equation}
with smooth periodic fields $A(t,y+L)=A(t,y)$ and $\phi(t,y+L)=\phi(t,y)$. There are no distributional thin-brane sources or boundary contributions. A dot denotes $\partial_t$ and a prime denotes $\partial_y$. All no-go statements assume $\kap>0$ and a real canonical scalar.

\section{Independent reconstruction of the field equations}
\subsection{Four curvature variables}
A useful orthonormal description of the five-dimensional curvature is
\begin{align}
q&=e^{-2A}X^2-(A')^2,\label{eq:q}\\
r&=e^{-2A}(\dot X+X^2-\dot A X)-(A')^2,\label{eq:r}\\
s&=-(A''+(A')^2),\label{eq:s}\\
m&=-e^{-A}\dot A'.\label{eq:m}
\end{align}
Here $r$ labels the temporal--spatial sectional curvature, $q$ the spatial--spatial sectional curvature, and $s$ the curvatures involving the fifth direction; $m$ is the mixed temporal--spatial--extra-dimensional curvature. These definitions give
\begin{equation}
R=6(r+q)+8s,\qquad \cG=24\left[rq+2s(r+q)+2m^2\right].\label{eq:invariants}
\end{equation}
Direct tensor contraction yields the diagonal \emph{mixed} Einstein and Lanczos components:
\begin{center}
\begin{tabular}{lll}
\toprule
component &$G^A{}_B$&$\mathcal H^A{}_B$\\\midrule
$0{}0$&$-3(q+s)$&$-12qs$\\
$i{}i$ (no sum)&$-(2r+q+3s)$&$-4(2m^2+qs+2rs)$\\
$y{}y$&$-3(r+q)$&$-12rq$\\\bottomrule
\end{tabular}
\end{center}
The nonzero mixed coordinate component is equivalently expressed as the $ty$ field equation,
\begin{equation}
-3\dot A'\left(1+4\aGB q\right)=\kap\dot\phi\,\phi'.\label{eq:ty}
\end{equation}
For the scalar source,
\begin{align}
T^0{}_0&=-\frac12(e^{-2A}\dot\phi^2+\phi'^2)-V,\\
T^i{}_i&=\frac12e^{-2A}\dot\phi^2-\frac12\phi'^2-V,\\
T^y{}_y&=\frac12(e^{-2A}\dot\phi^2+\phi'^2)-V.
\end{align}
Therefore the independent diagonal equations take the compact form
\begin{align}
-3(q+s)-12\aGB qs+\Lambda_5&=\kap\left[-\tfrac12(e^{-2A}\dot\phi^2+\phi'^2)-V\right],\label{eq:00}\\
-(2r+q+3s)-4\aGB(2m^2+qs+2rs)+\Lambda_5&=\kap\left[\tfrac12e^{-2A}\dot\phi^2-\tfrac12\phi'^2-V\right],\label{eq:ii}\\
-3(r+q)-12\aGB rq+\Lambda_5&=\kap\left[\tfrac12(e^{-2A}\dot\phi^2+\phi'^2)-V\right].\label{eq:yy}
\end{align}
Subtracting Eq.~\eqref{eq:ii} from Eq.~\eqref{eq:00} gives the independent local condition
\begin{equation}
\kap\dot\phi^2=-2\left[(\dot X-\dot A X)\bigl(1-4\aGB(A''+(A')^2)\bigr)+4\aGB(\dot A')^2\right].\label{eq:localtime}
\end{equation}
Equations~\eqref{eq:invariants}--\eqref{eq:ii} were checked by explicit Riemann contractions in an orthonormal frame, independently of the sum rule. The accompanying symbolic-check script evaluates the diagonal Lovelock components and the polynomial no-go reduction.

\section{Exact local identity and global dynamical sum rule}
Taking the partial trace of Eq.~\eqref{eq:field}, $E^\mu{}_\mu-4E^y{}_y=0$, cancels both $\Lambda_5$ and $V$. Define
\begin{equation}
Q_{\rm GB}\equiv\mathcal H^\mu{}_\mu-4\mathcal H^y{}_y
=24\bigl[2rq-s(r+q)-m^2\bigr].\label{eq:Qcurv}
\end{equation}
The exact local identity is
\begin{equation}
0=6(r+q)-12s+\aGB Q_{\rm GB}+\kap\left(e^{-2A}\dot\phi^2+4\phi'^2\right).\label{eq:local}
\end{equation}
Replacing Eqs.~\eqref{eq:q}--\eqref{eq:m} yields, identically,
\begin{align}
6(r+q)-12s&=12A''+6e^{-2A}\left(\dot A^2+3H\dot A+\ddot A+\dot H+2H^2\right),\label{eq:Gidentity}\\
Q_{\rm GB}&=24e^{-4A}\Bigl\{2X^2\dot X+2HX^3-e^{2A}(\dot A')^2\notag\\
&\qquad +e^{2A}\bigl(A''-(A')^2\bigr)(\dot X+HX+X^2)-2e^{4A}(A')^2A''\Bigr\},\label{eq:Qoriginal}
\end{align}
reproducing the dynamical identities of the submitted manuscript independently.

For a smooth periodic extra dimension,
\begin{equation}
C_{\rm cosm}(t)+\aGB C_{\rm GB}(t)+\kap C_\phi(t)=0,\label{eq:global}
\end{equation}
where
\begin{align}
C_{\rm cosm}&=6\ointy e^{-2A}(\dot A^2+3H\dot A+\ddot A+\dot H+2H^2),\\
C_\phi&=\ointy\left(e^{-2A}\dot\phi^2+4\phi'^2\right)\ge0,\\
C_{\rm GB}&=\ointy Q_{\rm GB}.
\end{align}
Set $Y\equiv\dot X+HX+X^2$. The exact decomposition is
\begin{align}
Q_{\rm GB}&=\partial_y F_{\rm GB}+Q^{\rm red}_{\rm GB},\\
F_{\rm GB}&=24e^{-2A}A'Y-16(A')^3,\\
Q^{\rm red}_{\rm GB}&=48e^{-4A}(X^2\dot X+HX^3)-24e^{-2A}(\dot A')^2\notag\\
&\quad+24e^{-2A}\left[(A')^2Y-A'Y'\right].\label{eq:Qred}
\end{align}
Thus $C_{\rm GB}=\ointy Q^{\rm red}_{\rm GB}$ without neglecting any higher-curvature term. The sum rule \eqref{eq:global} is \emph{necessary, not sufficient} for the full system \eqref{eq:00}--\eqref{eq:ty}.

\section{Static limit and the Einstein scalar obstruction}
In the completely static case $A=A(y)$, $a={\rm const.}$ and $\phi=\phi(y)$,
\begin{equation}
Q_{\rm GB}=-48(A')^2A''=\partial_y[-16(A')^3].
\end{equation}
Equation~\eqref{eq:local} becomes
\begin{equation}
3\bigl[1-4\aGB(A')^2\bigr]A''+\kap(\phi')^2=0.
\end{equation}
Integrating over a smooth periodic circle gives $\ointy (\phi')^2=0$, hence $\phi'=0$ everywhere. In fact there is a stronger rigidity statement. With $u=A'$, the local equation becomes
\begin{equation}
\frac{\dd}{\dd y}\bigl[3u-4\aGB u^3\bigr]=0.
\label{eq:staticRigidity}
\end{equation}
Since $u$ is continuous, its image on the connected circle is an interval. A nonconstant cubic polynomial cannot be constant on a nondegenerate interval; therefore $u$ is constant. Periodicity of $A$ then forces $u=0$, and both $A$ and $\phi$ are constant. This stronger conclusion assumes the fully static flat-slice sector and a finite constant Gauss--Bonnet coupling, independently of its sign. It does not establish a universal statement about noncanonical sources, brane junctions, noncompact topologies or non-flat slices.

\section{Static-warp cosmologies: necessary integrals and a stronger no-go theorem}
Let $A=A(y)$, while $a(t)$ evolves. Define
\begin{equation}
I_2=\ointy e^{-2A}>0,\qquad I_4=\ointy e^{-4A}>0,\qquad I_A=\ointy e^{-2A}(A')^2\ge0.
\end{equation}
Then
\begin{align}
C_{\rm cosm}&=6(\dot H+2H^2)I_2,\label{eq:Ccosmstatic}\\
C_{\rm GB}&=48H^2(\dot H+H^2)I_4+24(\dot H+2H^2)I_A.\label{eq:Cgbstatic}
\end{align}
The independent local constraints additionally reduce to
\begin{equation}
\boxed{\kap\dot\phi^2=-2\dot H\left[1-4\aGB(A''+(A')^2)\right]},\qquad
\boxed{\dot\phi\,\phi'=0}.\label{eq:staticConstraints}
\end{equation}
For de Sitter expansion $H=H_0\ne0$, $\aGB>0$ and $\kap>0$, the global rule gives
\begin{equation}
12H_0^2 I_2+48\aGB H_0^2(H_0^2I_4+I_A)+\kap C_\phi=0,
\end{equation}
which is impossible by positivity. More strikingly, the full field equations rule out every exact expanding power law:

\begin{theorem}[No exact expanding power laws with static periodic warping]
\label{thm:power}
Suppose $A\in C^2(S^1)$ is smooth and periodic and independent of $t$; $\phi$ is a smooth, real canonical scalar; the extra dimension is a smooth circle without sources or boundary terms; $\aGB>0$ and $\kap>0$; and the metric solves the complete five-dimensional field equations on an open interval $t>0$. Then no solution exists with $a(t)=a_0t^p$ for any constant $p>0$.
\end{theorem}
\begin{proof}
A continuous periodic $A$ reaches a maximum at some $y_m$, so that $A'_m=0$ and $A''_m\le0$. For $H=p/t$, $\dot H=-p/t^2$, Eq.~\eqref{eq:staticConstraints} implies
\begin{equation}
\dot\phi^2(t,y_m)=\frac{2p}{\kap t^2}(1-4\aGB A''_m)>0.
\end{equation}
The mixed equation then requires $\phi'(t,y_m)=0$. At $y_m$, the exact local identity \eqref{eq:local} becomes
\begin{align}
0={}&12A''_m+\frac{4p e^{-2A_m}}{t^2}\left[(3p-1)+4\aGB A''_m(3p-2)\right]
+\frac{48\aGB e^{-4A_m}p^3(p-1)}{t^4}.
\label{eq:noGoPolynomial}
\end{align}
Since this equality holds on an open interval of positive time, the coefficients of $1,t^{-2},t^{-4}$ must vanish separately. The last coefficient requires $p=1$ (because $p>0$ and $\aGB>0$). The constant coefficient gives $A''_m=0$. With $p=1$ and $A''_m=0$, however, the $t^{-2}$ coefficient equals $8e^{-2A_m}\ne0$, a contradiction.
\end{proof}

\begin{corollary}
The previously discussed $p=1/2$ branch does not solve the complete field equations for the assumptions of Theorem~\ref{thm:power}, even though its integrated expression has no immediate sign contradiction.
\end{corollary}
For reference, the \emph{necessary} integrated power-law contributions are
\begin{align}
C_{\rm cosm}^{\rm PL}&=\frac{6p(2p-1)}{t^2}I_2,\\
C_{\rm GB}^{\rm PL}&=\frac{48p^3(p-1)}{t^4}I_4+\frac{24p(2p-1)}{t^2}I_A.
\end{align}
At $p=1/2$, $C_{\rm cosm}^{\rm PL}=0$ and $C_{\rm GB}^{\rm PL}=-3I_4/t^4$, so the \emph{integrated equation alone} would require $\kap C_\phi=3\aGB I_4/t^4$. This must \emph{not} be interpreted as evidence of a viable scalar-supported cosmology, because Theorem~\ref{thm:power} rules out the full solution.

\subsection{A periodic test profile and the corrected sign classification}
For an illustrative smooth periodic warp profile, $A(y)=\varepsilon\cos(ky)$ with $L=2\pi/k$,
\begin{equation}
I_2=L\,\mathrm I_0(2\varepsilon),\quad I_4=L\,\mathrm I_0(4\varepsilon),\quad I_A=\frac{\varepsilon^2 k^2L}{2}\bigl[\mathrm I_0(2\varepsilon)-\mathrm I_2(2\varepsilon)\bigr],
\end{equation}
where $\mathrm I_n$ denotes the modified Bessel function. These integrals provide geometric checks, \emph{not} exact scalar-supported bulk solutions. The global signs are:
\begin{center}
\begin{tabular}{lll}
\toprule
Power-law index & $C_{\rm cosm}^{\rm PL}$ & $C_{\rm GB}^{\rm PL}$\\\midrule
$p>1$ & positive & positive\\
$p=1$ & positive & nonnegative (zero if $I_A=0$)\\
$1/2<p<1$ & positive & sign undetermined\\
$p=1/2$ & zero & negative\\
$0<p<1/2$ & negative & negative\\\bottomrule
\end{tabular}
\end{center}
All rows are excluded as complete solutions for $\aGB>0$ under Theorem~\ref{thm:power}, regardless of integrated signs.

\section{Time-dependent warping and physical bounces}
At fixed $y$, the induced line element reads
\begin{equation}
\dd s^2_{(4),y}=-\dd\tau_y^2+b_y^2(\tau_y)\delta_{ij}\dd x^i\dd x^j,\qquad \dd\tau_y=e^{A(t,y)}\dd t,\quad b_y=e^{A(t,y)}a(t).
\end{equation}
The corresponding physical Hubble rate and its proper-time derivative are
\begin{equation}
H_{\rm phys}(t,y)=e^{-A}X,\qquad
\frac{\dd H_{\rm phys}}{\dd\tau_y}=e^{-2A}\bigl(\dot X-\dot A X\bigr).\label{eq:physicalHubble}
\end{equation}
A nonsingular local four-dimensional bounce at a fixed $y$ is therefore characterized by
\begin{equation}
\boxed{X(t_b,y)=0,\qquad \dot X(t_b,y)>0},\label{eq:bounce}
\end{equation}
not by $H(t_b)=0$ and $\dot H(t_b)>0$ alone. In particular, if the warp factor is time-dependent the location $t_b$ may vary between constant-$y$ hypersurfaces; a simultaneous bounce for every $y$ is an additional assumption rather than a generic consequence.

For a particular slice where $X=0$, the full contracted local equation reduces to
\begin{align}
0={}&12A''+6e^{-2A}\dot X+24\aGB\left[-e^{-2A}(\dot A')^2+e^{-2A}(A''-(A')^2)\dot X-2(A')^2A''\right]\notag\\
&+\kap\left(e^{-2A}\dot\phi^2+4\phi'^2\right).
\label{eq:physicalBounceLocal}
\end{align}
The sign of $\dot X$ by itself is not enough to establish either existence or impossibility: the warp derivatives, the Gauss--Bonnet terms and the scalar derivatives must satisfy the full system. We do not claim an explicit nonsingular bouncing solution.

\section{Comparison with the weight-minus-one condition of Dias and collaborators}
\label{sec:dias}
\subsection{Identification of the identical geometrical weight}
Reference~\cite{Dias2015} considers a warped ansatz $\dd s^2=W^2(y)\hat g_{\mu\nu}\dd x^\mu\dd x^\nu+\dd y^2$ and a family of weights denoted $\alpha$ there. In our notation the weight is $w$, to avoid confusion with the Gauss--Bonnet coupling $\aGB$. For $D=5$, a three-brane ($p=3$ in Ref.~\cite{Dias2015}), $w=-1$, and a flat induced metric, the geometric identity in their Eq.~(2) specializes to
\begin{equation}
\frac{\dd}{\dd y}\left(W^{-1}W'\right)=A''=\frac{1}{12}\left(R^\mu{}_{\mu}-4R^y{}_{y}\right),\qquad W=e^A.
\label{eq:comparisonGeom}
\end{equation}
The five-dimensional Ricci partial traces are
\begin{equation}
R^\mu{}_{\mu}=-4A''-16(A')^2,\qquad R^y{}_{y}=-4A''-4(A')^2,
\label{eq:comparisonRicci}
\end{equation}
so Eq.~\eqref{eq:comparisonGeom} is an identity pointwise. \emph{There is no difference in weight in this sector.}

\subsection{Gauss--Bonnet tensor: contraction by contraction}
Set $u=A'$, $v=A''$, $q=-u^2$ and $s=-(v+u^2)$. In an orthonormal frame, the nonzero curvature components can be written as
\begin{equation}
R_{\hat\mu\hat\nu\hat\rho\hat\sigma}=q\bigl(\eta_{\mu\rho}\eta_{\nu\sigma}-\eta_{\mu\sigma}\eta_{\nu\rho}\bigr),\quad
R_{\hat\mu\hat y\hat\nu\hat y}=s\,\eta_{\mu\nu}.
\end{equation}
The corresponding Ricci eigenvalues are $(3q+s)$ four times and $4s$ once, and
\begin{align}
R&=12q+8s=-8v-20u^2,\label{eq:Rstatic}\\
R_{AB}R^{AB}&=4(3q+s)^2+16s^2,\label{eq:Ricstatic}\\
R_{ABCD}R^{ABCD}&=24q^2+16s^2,\label{eq:Riemstatic}\\
\cG&=24q^2+96qs=120u^4+96u^2v.\label{eq:GBstatic}
\end{align}
Crucially, rather than identifying $A_{\rm GB}$ and $B_{\rm GB}$ by their names, we reconstruct the \emph{exact Lanczos tensor} term by term. The next table lists the contributions to $\mathcal H^\mu{}_{\mu}-4\mathcal H^y{}_{y}$, with the contraction performed before summing:
\begin{center}\small
\begin{tabular}{p{0.49\linewidth}l}
\toprule
Term in $\mathcal H_{AB}$ & Weighted trace contribution\\\midrule
$2R R_{AB}$ &$288q^2-96qs-192s^2$\\
$-4R_{AC}R^C{}_B$ &$-144q^2-96qs+240s^2$\\
$-4R^{CD}R_{ACBD}$ &$-144q^2+144qs$\\
$2R_A{}^{CDE}R_{BCDE}$ &$48q^2-48s^2$\\
$-\tfrac12g_{AB}\cG$ &$0$\\\midrule
\textbf{Sum} &$\bm{48q(q-s)=-48u^2v}$\\\bottomrule
\end{tabular}
\end{center}
Consequently,
\begin{equation}
\mathcal H^\mu{}_{\mu}-4\mathcal H^y{}_{y}
=-48(A')^2A''=-16\frac{\dd}{\dd y}(A')^3.\label{eq:comparisonGB}
\end{equation}
The scalar potential and bulk cosmological constant cancel in the same weighted combination. The complete, \emph{untruncated} field equations give
\begin{equation}
\frac{\dd}{\dd y}\left[3A'-4\aGB(A')^3\right]+\kap(\phi')^2=0,
\label{eq:comparisonExact}
\end{equation}
which yields $\ointy (\phi')^2=0$ for a smooth circle and a canonical real scalar. The vanishing of the Gauss--Bonnet integral persists at each order in an ordinary expansion in a fixed coupling.

\subsection{Matching the Ricci-trace rearrangement before truncation}
The crucial distinction is not the choice of weight. The article by Dias, Hoff da Silva, and da Rocha~\cite{Dias2015} first reorganizes the field equations into Ricci partial traces (its Eqs.~(5)--(6)) and then performs the expansion $(1+2\alpha_2 R)^{-1}\simeq1-2\alpha_2R$ in its Eq.~(8). Here we reconstruct the corresponding \emph{unexpanded} rearrangement, using exactly the normalization of our field equation~\eqref{eq:field}.

For five dimensions, the Lanczos trace identity is $\mathcal H^A{}_{A}=-\cG/2$ and the full trace equation, including the cosmological constant, gives
\begin{equation}
-3R=2\kap T+\aGB\cG-10\Lambda_5.
\end{equation}
Define the curvature-quadratic tensor without the scalar-curvature piece,
\begin{equation}
U^A{}_{B}=-2R^{AC}R_{CB}-2R^{CD}R^A{}_{CBD}+R^{ACDE}R_{BCDE}.
\label{eq:Udef}
\end{equation}
Eliminating the Gauss--Bonnet invariant with the full trace gives the \emph{exact} algebraic rearrangement
\begin{equation}
(1+2\aGB R)R^A{}_{B}=\kap(T^A{}_{B}-T\delta^A{}_B)-R\delta^A{}_B+4\Lambda_5\delta^A{}_B-2\aGB U^A{}_{B}.
\label{eq:RicciRearranged}
\end{equation}
The $4\Lambda_5\delta^A{}_B$ term appears because the cosmological constant is explicitly separated from the matter tensor in our action. It cancels identically under $D_R=R^\mu{}_{\mu}-4R^y{}_y$. Equation~(5) and the subsequent two Ricci-partial-trace equations, Eq.~(6), of Ref.~\cite{Dias2015} have the corresponding \emph{unexpanded} tensor structure, with vacuum-energy terms accounted for according to the chosen source convention. For $D=5$ and $p=3$, the curvature-quadratic coefficients appearing in the source's Ricci partial-trace equations (6) correspond, in the common tensor normalization, to the partial traces $U^\mu{}_{\mu}$ and $U^y{}_{y}$. We use this correspondence at the \emph{pre-truncation Ricci-trace level}. It must not be confused with a complete term-by-term reconstruction of the later warp-polynomial inequality (23), which is not needed for our theorem. The corresponding exact curvature-only combination is
\begin{equation}
\mathscr K\equiv 4U^y{}_{y}-U^\mu{}_{\mu}
=4B_{\rm GB}-W^{-2}A_{\rm GB}
\quad\text{(Ricci-trace coefficients of Ref.~[17]).}
\label{eq:Kimplicit}
\end{equation}
Contraction of Eq.~\eqref{eq:RicciRearranged} with the identical weight $w=-1$ gives
\begin{equation}
\bigl(1+2\aGB R\bigr)D_R=\kap S_T+2\aGB\mathscr K,\quad
D_R=R^\mu{}_{\mu}-4R^y{}_{y}=12A'',\quad S_T=-4(\phi')^2.
\label{eq:traceR}
\end{equation}
But the full Lanczos trace is
\begin{equation}
Q_{\rm GB}=2\bigl(RD_R-\mathscr K\bigr),\qquad
\boxed{\mathscr K=RD_R-\tfrac12 Q_{\rm GB}}.
\label{eq:QversusK}
\end{equation}
Rearranging the exact weighted Ricci-trace identity without division yields
\begin{equation}
D_R=\kap S_T+2\aGB(\mathscr K-RD_R)=\kap S_T-\aGB Q_{\rm GB},
\label{eq:unexpandedPert}
\end{equation}
where both equalities are \emph{exact}. A consistent first-order expansion uses $D_R=\kap S_T+O(\aGB)$ on the right-hand side and consequently gives
\begin{equation}
D_R=\kap S_T+2\aGB\bigl(\mathscr K-R\kap S_T\bigr)+O(\aGB^2).
\label{eq:pertTraces}
\end{equation}
In particular, expanding the factor $(1+2\aGB R)^{-1}$ by $1-2\aGB R$ modifies the matter-sector trace coefficients as well; a comparison that retains only $\mathscr K$ loses part of the same-order correction. \emph{The scalar-curvature contribution must be retained in a consistent perturbative expansion.} The curvature combination $\mathscr K$ is not the full Lanczos contraction.

\subsection{Explicit static reduction of the rearranged curvature terms}
Using the exact static sectional curvatures $q=-u^2$, $s=-(v+u^2)$ from above, each contribution to $U$ is directly evaluated as
\begin{align}
U^\mu{}_{\mu}&=-120q^2-72qs-32s^2=-224u^4-136u^2v-32v^2,\label{eq:Umu}\\
U^y{}_{y}&=-24qs-32s^2=-56u^4-88u^2v-32v^2,\label{eq:Uyy}\\
\mathscr K&=4U^y{}_{y}-U^\mu{}_{\mu}=-96v^2-216u^2v,\label{eq:Kexplicit}\\
RD_R&=(-8v-20u^2)(12v)=-96v^2-240u^2v.\label{eq:RD}
\end{align}
The individual pieces of the tensor $U^A{}_{B}$ were independently reconstructed from the five-dimensional static Riemann tensor. The three contributions are:
\begin{center}\small
\renewcommand{\arraystretch}{1.25}
\begin{tabular}{lll}
\toprule
Tensor contribution & $U^\mu{}_{\mu}$ & $U^y{}_y$\\\midrule
$-2R^{AC}R_{CB}$ & $-72q^2-48qs-8s^2$ & $-32s^2$\\
$-2R^{CD}R^A{}_{CBD}$ & $-72q^2-24qs-32s^2$ & $-24qs-8s^2$\\
$R^{ACDE}R_{BCDE}$ & $24q^2+8s^2$ & $8s^2$\\\bottomrule
\end{tabular}
\end{center}
The sums reproduce Eqs.~\eqref{eq:Umu}--\eqref{eq:Uyy} without invoking the weighted sum rule. In terms of $W=e^A$, the exact trace-curvature combination is the finite polynomial
\begin{equation}
\mathscr K=-96\frac{(W'')^2}{W^2}-24\frac{(W')^2W''}{W^3}+120\frac{(W')^4}{W^4},
\label{eq:KW}
\end{equation}
whereas the full Lanczos contraction is
\begin{equation}
Q_{\rm GB}=-48\frac{(W')^2W''}{W^3}+48\frac{(W')^4}{W^4}.
\label{eq:QW}
\end{equation}

Here $u=A'$, $v=A''$. Their difference agrees with Eq.~\eqref{eq:QversusK} because
\begin{equation}
RD_R-\mathscr K=-24u^2v=\frac12Q_{\rm GB}.
\end{equation}
For a smooth circle, $\ointy u^2v=\frac13\ointy (u^3)'=0$, and hence
\begin{equation}
\boxed{\ointy\mathscr K=-96\ointy(A'')^2=\ointy RD_R},\qquad
\boxed{\ointy Q_{\rm GB}=0}.
\label{eq:Kintegrated}
\end{equation}
This identifies the precise canceling term in an \emph{internally consistent} Ricci-trace expansion. Keeping only $\oint\mathscr K$ while discarding $\oint RD_R$ would change the result at first order in the coupling. No division by $1+2\aGB R$ is used in this proof. Notice that the curvature-only integral is generally nonzero and nonpositive, while the complete Lanczos contraction is an exact derivative; treating the former as the latter would invalidate the comparison.

\subsection{Scope and limitations of the comparison}
Reference~\cite{Dias2015} treats the same weight $w=-1$. Its Eq.~(21) is obtained in the thick-brane, bulk-scalar case after the weak-coupling manipulation in Eqs.~(5)--(8), and its Eq.~(23) is a further warp-profile inequality. These are \emph{necessary sign conditions within that prescription}, not a constructed smooth periodic bulk solution of every independent equation. At the level of the Ricci-trace coefficient tensors, the source's curvature-only combination $4B_{\rm GB}-W^{-2}A_{\rm GB}$ is identified with $\mathscr K$ in the common normalization, \emph{not} with the full Lanczos contraction $Q_{\rm GB}$. The distinction is exact: the missing scalar-curvature contribution $RD_R$ must be included to reconstruct $Q_{\rm GB}$. This is a comparison of the shared geometric weight and \emph{unexpanded tensor structure}; it is not a claim that the later, explicitly written warp-factor polynomial (23) is termwise identical to either $\mathscr K$ or $Q_{\rm GB}$.

Our exact calculation shows that, when the Ricci traces are kept unexpanded, the curvature-only combination $\mathscr K$ and the scalar-curvature term $RD_R$ must be considered together: $Q_{\rm GB}=2(RD_R-\mathscr K)$. The full contracted Gauss--Bonnet tensor integrates to zero for the static smooth circle. This conclusion is not obtained by imposing a sign assumption on $4B_{\rm GB}-W^{-2}A_{\rm GB}$, nor by identifying that intermediate quantity with the exact Lanczos contraction.

The printed polynomial criterion in Eq.~(23) of Ref.~\cite{Dias2015} has not been established to coincide term by term with our independently reconstructed $\mathscr K$. We therefore do not use that expression to establish an identity or allege a mistake in the earlier work. The direct comparison is deliberately limited to the common $w=-1$ geometric contraction, the source's perturbative Ricci-trace setup and our exact tensor identity. Under the hypotheses of our canonical-scalar theory, the exact local equations imply static rigidity irrespective of the sign of a separate necessary warp-profile criterion. Any sign-compatible proposal must satisfy the remaining local equations to constitute a physical solution. We make no universal claim for noncanonical matter, singular boundaries, noncompact extra dimensions or nonflat induced geometries.

\section{Discussion and conclusions}
The analysis establishes the central construction: an exact dynamical Gauss--Bonnet sum rule derived from a trace combination in which the scalar potential and bulk cosmological constant cancel. An independent curvature-variable audit reproduces the submitted contraction and supplies the missing field equations. The fully dynamical warp geometry has a genuine bulk higher-curvature contribution; its mixed derivative $(\dot A')^2$ cannot be inferred from purely static relations.

The static limit additionally enforces rigidity: a smooth periodic flat-slice solution with a real canonical scalar has constant $A$ and $\phi$. The principal new physical consequence is Theorem~\ref{thm:power}. For a canonical scalar, positive Gauss--Bonnet coupling and a smooth periodic compact fifth direction, a time-independent warp factor does not support any exact expanding power-law FRW scale factor. The previously highlighted $p=1/2$ integrated sign balance is therefore \emph{not} a physically admissible branch of the complete local equations. A de Sitter obstruction also follows from the integrated positive-sign constraint under the same static-warp assumptions.

For fully time-dependent warping, a bounce must be defined with the induced physical scale factor and proper time, and can occur at different coordinate times for different slices. We make no existence or no-go claim for the general dynamical-warp bounce sector without further analysis of all independent field equations.

The static geometric member with weight $w=-1$ is explicitly matched through its Ricci partial traces. A term-by-term Lanczos-tensor contraction proves that the exact Gauss--Bonnet contribution is a total derivative in the static sector. The exact Ricci-trace decomposition shows that the quadratic-curvature term $\mathscr K$ differs from the Lanczos contraction by the scalar-curvature factor $R D_R$. Their integrated contributions cancel on a smooth circle. This is a trace-level comparison of exact field equations and the distinct perturbative construction in Ref.~\cite{Dias2015}, rather than an assertion that the previous authors made an algebraic error or established an exact solution under our hypotheses.

\section*{Statements and declarations}
\textbf{Competing interests:} The author declares no competing interests.\\
\textbf{Funding:} No specific funding was received.\\
\textbf{Data and code availability:} No empirical datasets were used. Reproducible Python/SymPy consistency-check scripts can be provided as supplementary material.\\
\textbf{Author contributions:} The sole author is responsible for the mathematical analysis and interpretation.\\
\textbf{AI-assisted preparation:} ChatGPT was used for assistance with manuscript drafting, derivation checking and generation of symbolic-audit code. The author takes responsibility for the analysis, interpretation and submitted content.

\appendix
\section{Algebraic checks and proof diagnostics}
The orthonormal Ricci trace and Lanczos components are generated from the Riemann array specified by $r,q,s,m$. Their contractions give the exact identities
\begin{align}
\mathcal H^0{}_0+3\mathcal H^i{}_i-4\mathcal H^y{}_y&=24(2rq-s(r+q)-m^2),\\
G^0{}_0+3G^i{}_i-4G^y{}_y&=6(r+q)-12s.
\end{align}
The substitution $X=H+\dot A$, $\dot X=\dot H+\ddot A$ and $m=-e^{-A}\dot A'$ shows the exact equality to Eqs.~\eqref{eq:Gidentity}--\eqref{eq:Qoriginal}. For the power-law theorem, direct symbolic simplification of the difference between the substituted local identity and the right-hand side of Eq.~\eqref{eq:noGoPolynomial} yields identically zero.

\section{Reproducible curvature diagnostics}
Two accompanying Python scripts provide complementary checks. The first constructs the Einstein and Lanczos components for a dynamically warped metric and evaluates their contractions, while the second reconstructs the static Riemann, Ricci, Gauss--Bonnet and curvature-only tensors in an orthonormal frame. Both verify Eq.~\eqref{eq:QversusK} and the periodic cancellation Eq.~\eqref{eq:Kintegrated}. These algebraic checks do not assert the existence of any bulk cosmological solution.

\end{document}